\documentclass[twocolumn,10pt,superscriptaddress]{revtex4}

\usepackage{graphicx}
\usepackage{amssymb}
\usepackage{amsmath}
\usepackage{upgreek}				% For upright greek characters

\begin{document}
\title{Phonon dispersion curves of two-dimensional colloidal crystals:
  on the wavelength dependence of friction}
  
\author{J. Baumgartl}
  \affiliation{2.Physikalisches Institut, Universit\"at Stuttgart,
  Stuttgart, Germany}
\author{J. Dietrich} 
  \affiliation{2.Physikalisches Institut, Universit\"at Stuttgart,
  Stuttgart, Germany}
\author{J. Dobnikar} 
\affiliation{Jozef Stefan Institute, Jamova 39,
  1000 Ljubljana, Slovenia} 
\author{C. Bechinger}
  \affiliation{2.Physikalisches Institut, Universit\"at Stuttgart,
  Stuttgart, Germany} 
  \affiliation{Max-Planck-Institut f\"ur
  Metallforschung, Stuttgart, Germany}  
  \author{H.H. von Gr\"unberg}
\affiliation{Institut f\"ur Chemie, Karl-Franzens-Universit\"at,
  Graz, Austria} \date{\today}
  
\begin{abstract}
Digital video-microscopy measurements are reported of both elastic
bandstructures and overdamped phonon decay times in two-dimensional
colloidal crystals. Both quantities together allow to determine the
friction coefficients along various high symmetry directions in
$\vec{q}$-space. These coefficients contain valuable information on
the hydrodynamic forces acting between the colloidal particles. We
find Stokes-like friction for phonons near the edge of the first
Brillouin zone and vanishing friction coefficients for long wavelength
phonons. The effect of this wavelength dependence in real-space is
further investigated by simulating a crystal with constant friction
(Langevin simulation) and comparing experimentally measured and
simulated particle auto-correlation functions.
\end{abstract}

  
  \maketitle

\section{Introduction}

The lattice dynamics of crystals composed of colloidal particles is
entirely different from that of atoms. Being immersed in a viscous
solvent, colloids experience friction which strongly dampens their
motion. As a result, phonons in colloidal systems -- characterized by
the polarization index $j$ and the wave-vector $\vec{q}$ -- show an
overdamped dynamics and decay exponentially with decay times
$T(\vec{q}j)$. These times are related to the eigenvalues
$\lambda(\vec{q}j)$ of the dynamical matrix which are refered to as the
elastic bandstructure in the following.  These eigenvalues may be
pictured as the ``spring-constant'' associated with the phonon
$(\vec{q}j)$. In overdamped systems, the relation between
$T(\vec{q}j)$ and $\lambda(\vec{q}j)$ is the phonon dispersion curve
which replaces the relation between the frequency of the propagating
phonons and the bandstructure $\lambda(\vec{q}j)$ in atomic
systems. The phonon dispersion curve quantifies the response of a
crystal to perturbations of a given wavelength; as such it is {\em
the} central relation of every crystal, characterizing both its static
and dynamical properties.

Contrary to atomic systems, where the phonon dispersion curve is
entirely determined by interparticle forces, in colloidal systems
hydrodynamic interactions have to be considered. These interactions
arise when moving colloids exchange momentum through the
solvent. Because of their long-ranged nature these interactions are
difficult to treat theoretically, especially in crystals
\cite{hasimoto1959}. It has been demonstrated by Hurd et
al. \cite{hurd1982,hurd1985} that along certain high symmetry
directions within the colloidal crystal hydrodynamic interactions can
be taken into account through the wavelength dependence of the
friction coefficients $\gamma(\vec{q}j)$. These coefficients connect
the phonon decay times and the bandstructure such that in the limit
of strong damping ($\gamma(\vec{q}j) \gg \lambda(\vec{q}j)$) the
phonon dispersion relation becomes
$T(\vec{q}j)=\gamma(\vec{q}j)/\lambda(\vec{q}j)$. Hence, with the
wavelength dependence of the friction coefficients one obtains
detailed information also on the hydrodynamic forces acting within
colloidal crystals.

The purpose of this paper is to report measurements of colloidal
crystal dispersion curves of thermally excited overdamped phonons,
and, more specifically, measurements of the $\vec{q}$-dependence of
the friction coefficients involved. We will compare the experimental
data with Langevin simulations where the friction is taken into
account only by the constant Stokes friction. This comparison will
demonstrate what effect the wavelength dependence of the friction can
have on correlations in real space. The simulations also serve to
check for the consistency of the evaluation procedure, for finite size
effects as well as possible sampling errors.

Measuring $\vec{q}$-dependent friction coefficients in colloidal
crystals is experimentally demanding, and such experiments have -- to
our knowledge -- never been attempted. Such studies require the
simultaneous measurement of both the decay times and the
bandstructure.  While decay times are experimentally accessible with
dynamic light scattering techniques
\cite{hurd1982,derksen1992,hoppenbrouwers1998,cheng2000,tat2004} or
Brillouin scattering \cite{penciu2002,penciu2003}, the experimental
determination of $\lambda(\vec{q}j)$ requires real space positional
information. Due to the mesoscopic time- and length scales, such
information is accessible in colloidal systems with digital video
microscopy which allows to follow particle positions with a resolution
of a few nm at video rates as high as 25 Hz.

The literature on phonon dynamics in overdamped systems is still
manageable.  Of central importance to the present paper is the work of
Hurd et al. \cite{hurd1982} who starting from \cite{hasimoto1959} not
only present a theory of hydrodynamic interaction in colloidal
crystals but also performed photon-correlation spectroscopy
measurements to study the dispersion of lattice waves. Early studies
of single-particle dynamics within colloidal crystals employed
scattering techniques \cite{piazza1991,brands1999}, while the first
video-microscopy experiments relevant for our question were presented
by Bongers and Versmold \cite{bongers1995} who measured the particle
mean square displacement as a function of time, later theoretically
analysed within the framework of the Langevin model in
\cite{ohshima2001,ohshima2001b}.

Hydrodynamics within colloidal systems that are not in the crystalline
state, is the subject of a large number of papers. Digital video
microscopy has been used to investigate dynamic properties (dynamic
structure factor, the hydrodynamic function, hydrodynamic diffusion
coefficients) in colloidal suspensions confined between two parallel
glass plates \cite{carba1997,arauz01,arauz05}. Colloidal friction has
also been a major issue in recent optical tweezer measurements of the
hydrodynamic interaction between two colloidal particles
\cite{meiners1999,henderson2001,metzger2007}.  The time-independent
hydrodynamic forces as well as the cross-correlation function of the
colloid position have been interpreted in terms of two-body
hydrodynamic interactions, embodied by the Oseen tensor. However, a
two-body description of hydrodynamics in crystals is inadequat because
of the neglect of many-body effects which dominate the hydrodynamic
interaction in regular arrays of particles
\cite{hurd1982,hasimoto1959}.

When at long wavelengths the frictional force vanishes, damping can
become so weak that even in colloidal systems phonons should be able
to propagate. Hurd et al. predicted a switch to such propagating
modes for transverse modes but were not able to observe them, probably
because of the finite-cell geometry and hydrodynamic wall effects -- a
conclusion that has later been corroborated by Derksen et
al. \cite{derksen1992} who carefully reconsidered the Hurd
experiment. An alternative explanation offered by Felderhof and Jones
\cite{felderhof1987} in terms of additional damping through retarded
counterions was challenged by Derksen et al. \cite{derksen1992} and
Hoppenbrouwers et al. \cite{hoppenbrouwers1998}. Evidence for
overdamped transverse modes turning propagative at long wavelengths
has later been produced by Tata et al. \cite{tat2004} in finite size
millimeter crystals. We here completely ignore such propagating modes,
returning to a brief discussion of their possible existence only in
the last section.

The paper starts with a short presentation of the overdamped Langevin
model of a colloidal crystal, which forms the basis of our
simulations. Then, the experimental setup is described, followed by a
section in which we outline how bandstructures, friction coefficients
and correlation functions are obtained from particle
configurations. We proceed with the coefficients and the particle auto-correlation 
functions to show the differences between systems with and without wavelength-dependent
friction, i.e., the differences between simulation and experiment. The
main conclusions of the paper are formulated in the last section.

\section{Experimental}
\subsection{Simulation: The overdamped Langevin model of a colloidal crystal}

We consider a two-dimensional (2D) hexagonal crystal of $N$
colloidal spheres with diameter $\sigma$ immersed in an aqueous
electrolyte being characterized by the dielectric constant
$\epsilon$ and the inverse screening length $\kappa$. 
Particle positions at time $t$ are denoted by $\vec{x}_n(t)$ with
$n$ labeling a hexagonal lattice site $\vec{R}_n$. The position vector
decomposes into
$\vec{x}_n(t)=\vec{R}_n+\vec{u}_n(t)$ with
$\vec{u}_n(t)$ being the particle displacement from the
lattice site $\vec{R}_n$. The interaction of two colloids at
distance $r$ is given by the Yukawa potential,
\begin{equation}\label{eq1}
\Phi(r)=\frac{(Z_{eff}e)^{2}}{4\uppi\varepsilon_{0}
\varepsilon}\frac{\exp(\kappa\sigma)}{(1+\kappa\sigma/2)^2}
\frac{\exp(-\kappa{}r)}{r}
\end{equation}
where $Z_{eff}$ is the effective colloidal charge. Within the
framework of the harmonic approximation, the particles interact
through elastic forces characterized by the spring constant $k_{0}$,
given by
\begin{equation}\label{eq2}
k_{0}:=\bigg[\frac{d^{2}\Phi(r)}{dr^{2}}-
\frac{1}{r}\frac{d\Phi(r)}{dr}\bigg]_{r=a}
\end{equation}
where $a$ is the nearest-neighbor distance in the hexagonal crystal.

As explained in the next section, in our experiments the colloidal
crystal is stabilized by a commensurate hexagonal, light-induced
substrate created by three interfering laser beams. In the sample
plane, the electric fields of the linearly polarized laser beams are
given by
\begin{eqnarray}\label{eq2a}
\vec{E}_{j}(\vec{x}) & = &\vec{A}_{0}\exp(\mathrm{i}\vec{K}_{j}\vec{x})\quad j=1,2,3\\
\nonumber
\mbox{with }\vec{K}_{1}  &=&  K a\vec{e}_{y},\\
\nonumber
\vec{K}_{2}& = &K\big(-\frac{\sqrt{3}}{2}a
\vec{e}_{x}-\frac{a}{2}\vec{e}_{y}\big),\\
\vec{K}_{3}  &=&
K\big(\frac{\sqrt{3}}{2}a\vec{e}_{x}-\frac{a}{2}\vec{e}_{y}\big) \nonumber
\end{eqnarray}
where $\vec{e}_{x}=[1,0]$ and $\vec{e}_{y}=[0,1]$ are the basis
vectors of Cartesian coordinates, and $K$ is chosen to match
commensurate conditions. The interfering laser beams create the
substrate potential
\begin{eqnarray}\label{eq3}
\nonumber
\Phi_{ext}(\vec{x}) &=&
\frac{A_{0}^{2}}{2}\bigg( 9 - \bigg|
\sum_{i=1}^{3}\exp(\mathrm{i}\vec{K}_{i}\vec{x})\bigg|^{2} \bigg)\\
 &=& A_{0}^{2}\Bigg(3 - \sum_{i<j} \cos\big[\vec{K}_{ij}\vec{x}\big]\Bigg)
\end{eqnarray}
where all three vectors $\vec{K}_{ij} = \vec{K}_{i}- \vec{K}_{j}$
must have a length $K'=4 \uppi /(\sqrt{3}a)$ to ensure a fully
commensurate substrate. To adapt this potential to the harmonic
approximation used in our simulations, we expanded the cosine
function in eq.~(\ref{eq3}) for positions
$\vec{x}_n=\vec{R}_n+\vec{u}_n$ near a lattice site $\vec{R}_n$, and
finally obtain
\begin{equation}\label{eq3b}
  \Phi_{ext}(\vec{R}_n+\vec{u}_n) \approx
  \frac{A_{0}^{2} K'^2}{2}
\sum_{i<j} \big[\vec{K}'_{ij}\vec{u}_n\big]^2
\end{equation}
where $\vec{K}'_{ij}=\vec{K}_{ij}/K'$ and where use was made of
$\vec{K}_{ij}\cdot \vec{R}_n = 2 \uppi m$ with an integer $m$.  Thus, we
arrive at another spring constant
\begin{equation}\label{eq3c}
k_{1}:= A_{0}^{2}K'^2 = \bigg(\frac{A_0 4 \uppi}{\sqrt{3} a}\bigg)^2
\end{equation}
defining the strength of the springs with which every particle is
pinned to its lattice site $\vec{R}_n$.

Having introduced $k_0$ and $k_1$ we can now proceed to the model
equation describing the dynamics of the colloidal crystal. We here
adopt a simplified Langevin description of the lattice dynamics,
ignore hydrodynamic interactions and include just the Stokes
friction between the colloids and the solvent.  For more elaborate
lattice dynamic theories for colloidal crystals including
hydrodynamic interactions, see
\cite{hurd1982,felderhof1986,felderhof1987,schram1996,hofman1999}.

If $\vec{R}_{n}$ is an arbitrary reference site of the hexagonal
lattice with its six nearest neighbors
$\vec{R}_{1}=\vec{R}_{n}+a\vec{e}_{y}$,
$\vec{R}_{2}=\vec{R}_{n}-\sqrt{3}a/2\vec{e}_{x}+a/2\vec{e}_{y}$, and
so forth, the overdamped Langevin equation in the harmonic
approximation reads
\begin{align}
   \label{eq5}
   \nonumber
&\gamma\frac{\partial{} u_{n\alpha}(t)}{\partial{}t} \\
\nonumber
& +\frac{k_{0}}{a^2}
\sum_{m=1}^{6}\sum_{\beta=x,y} (R_{m\alpha}-R_{n\alpha})
(R_{m\beta}-R_{n\beta})\\
&\times[u_{m}(t)-u_{n}(t)]+
k_{1}u_{n\alpha}(t)+f_{n\alpha}(t)=0.
\end{align}
where we have restricted ourselves to nearest-neighbor interactions
and where an insignificant term $\sim \delta_{\alpha\beta}\Phi'(r)/r$
has been omitted. $\vec{f}_{n}(t)$ is a randomly fluctuating force
for which the following thermal averages must be satisfied
\begin{align}
\nonumber
\langle f_{n\alpha}(t) \rangle &= 0 \:,\\
 \langle
f_{n\alpha}(t) f_{n'\alpha'}(t+\tau)\rangle &= 2 \gamma k_{\mathrm{B}}T \delta(\tau)
\delta_{nn'} \delta_{\alpha\alpha'}
\end{align}
with $k_{\mathrm{B}}T$ being the thermal energy. In eq.~(\ref{eq5}), we have
introduced the Stokes friction coefficient $\gamma$ which measures how
strongly a quiescent fluid resists the motion of an isolated
sphere. Clearly, this is a rough approximation as the results of the
present paper will once more confirm: neither can a crystal particle
be considered isolated, nor is the surrounding fluid quiescent, but
will in fact be itself in a dynamical state due to the motion of all
other particles.

One can express the times in units of $\gamma \,\upmu \mathrm{m}^2/k_{\mathrm{B}}T$ and the
spring constant $k_0$ in terms of $k_{\mathrm{B}}T/\upmu \mathrm{m}^2$ such that the number of
independent input parameters to eq.~(\ref{eq5}) is reduced to just
three values. In our simulations we have chosen $k_0 \,\upmu \mathrm{m}^2 /k_{\mathrm{B}}T =
20$, $k_1/k_0 = 0.6$ and $a=4 \,\upmu \mathrm{m}$. While the simulations
have been performed with reduced units, the data will be presented in
normal physical units.

Finally, it must be mentioned that for technical reasons
a value of $k_0 \,\upmu \mathrm{m}^2 /k_{\mathrm{B}}T =20$ and above turned out to be the most
convenient choice. These values are well above the ones observed
in our experiments; however, the
band structure scales with $k_{0}$ as has been
shown in [25] and, therefore, our simulations
allow for a meaningful analysis of the
experimental results.

\subsection{Experiment: 2D colloidal crystals stabilized by light-induced substrates}

\begin{figure}[ht]
    \centering
\includegraphics[width=0.48\textwidth]{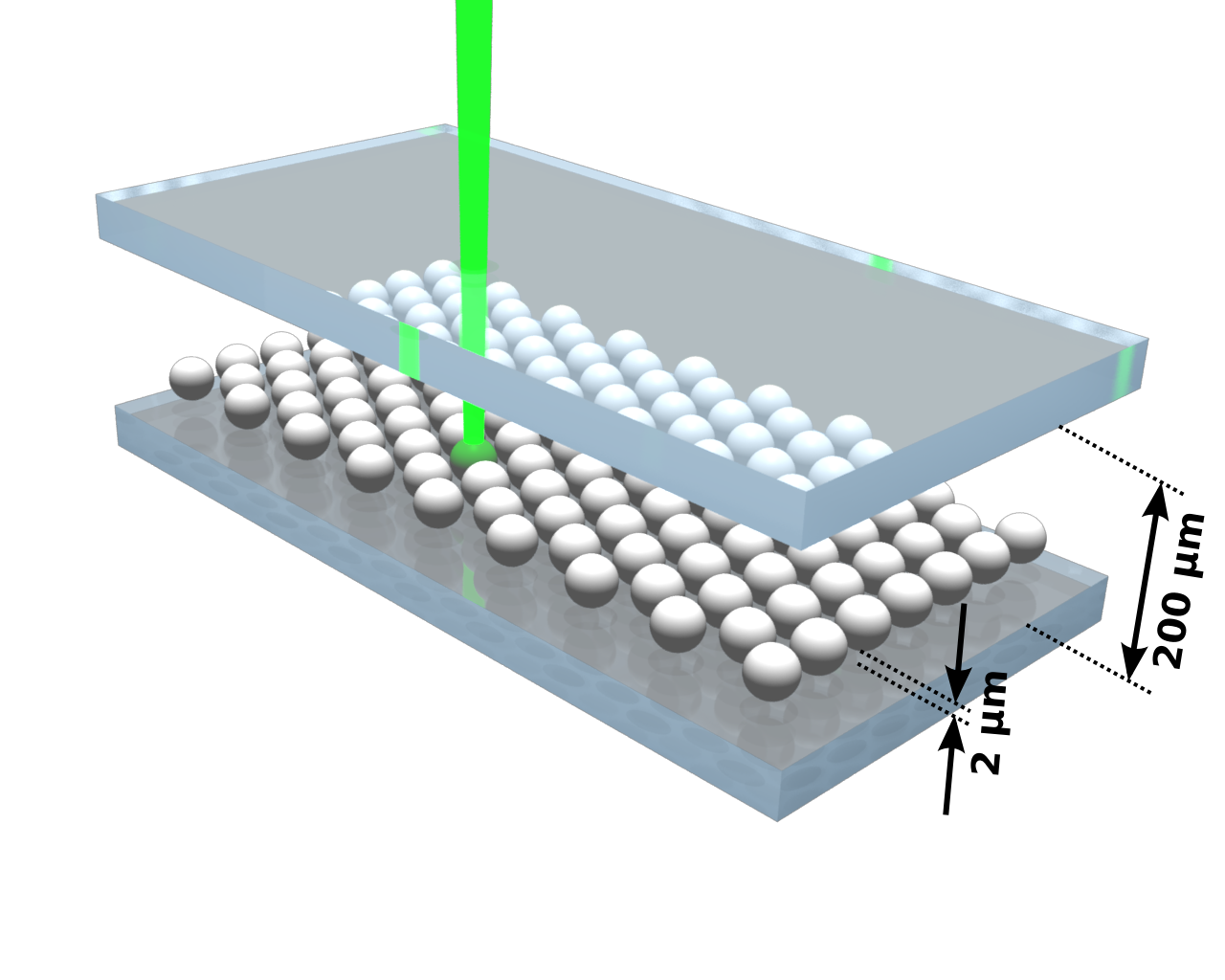}
\caption{Sketch of the sample cell containing the 2D crystal subject
  to a light-induced substrate (indicated for one particle) and, additionally, pressed down towards
  the bottom plate by an normal incident laser beam.}
    \label{fig1}
\end{figure}

Experiments were performed with an aqueous suspension of highly
charged polystyrene spheres with $\sigma = 2.4\,\upmu \mathrm{m}$ diameter and a
polydispersity below 4\%. The particles interact via a screened
Coulomb potential, eq.~(\ref{eq1}), with a renormalized surface charge
of about $Z_{eff} \approx 14000$ \cite{mariaPRL} and a screening
length $\kappa^{-1}\approx 458 \,\mathrm{nm}$, determined as described further
below. As sample cell we used a cuvette made of fused silica with
$200\,\upmu \mathrm{m}$ spacing between top and bottom plate which was connected to
a standard closed deionization circuit to maintain stable ionic
conditions during the measurements. An Argon-ion laser beam (wave
length = $488 \,\mathrm{nm}$) was scanned around the central region of the sample
to create a boundary box whose size could be continuously adjusted by
a pair of computer-controlled galvanostatically driven mirrors. This
allowed us to adjust precisely the particle area density
\cite{brunnerEPL}. At sufficiently high particle densities the
particles form a hexagonal 2D colloidal crystal ($a=6 \,\upmu \mathrm{m}$) close to
the bottom plate (see Fig.~\ref{fig1}). Out-of-plane fluctuations
are reduced by pushing the particles down towards the bottom plate
with a normal incident laser beam. Due to the interplay of the
electrostatic particle-wall repulsion and the light pressure from
above, this results in a mean particle-wall distance of about $2\,\upmu \mathrm{m}$
and normal fluctuations of an amplitude of less than $100\,\mathrm{nm}$. By
systematic variation of the light pressure we have verified that
within our parameters it does not influence the results discussed
below.

It is known that 2D crystals cannot have perfect long-ranged
translational order because low-$\vec{q}$ phonons destabilize the
system. This would considerably reduce the quality of our measured
particle configurations and the correlation functions generated from
it. We therefore stabilized the crystal by exposing it to the
perfectly hexagonal commensurate substrate potential from
eq.~(\ref{eq3}) being created from the interference of the three laser
beams in eq.~(\ref{eq2a}) ($P = 5\,\mathrm{W}$, frequency-doubled Nd:YVO4 laser,
wavelength $= 532 \,\mathrm{nm}$). It has been shown theoretically and
experimentally \cite{baumgartl07,mariaPRL} that if the light-induced
substrate is created from laser beams of the same intensity the
resulting bandstructure is identical to that of the free crystal,
except that it is shifted by $k_1$.

Particle positions and trajectories were determined for a 16 shell
hexagonal crystallite (N = 817 particles) which was part of a much
larger crystal comprising about 2000 to 3000 particles. 
All derived quantities were computed from
averages over 29500 images, using digital video microscopy at an
aquisition rate of 8 frames per second.

\subsection{Data processing: from configurations to derived quantities}

Simulation and experiment produce the same output: sets of particle
configurations. These will then be further processed to generate what
may be called ``derived quantities''. Both sets of configurations, the
computer-generated ones as well as the experimental ones, will be
processed in the same way.

It is a straightforward procedure to compute for a given
configuration the normal coordinates $Q(\vec{q}j,t)$ directly from
the measured displacements $\vec{u}_n(t)$
\cite{baumgartl07}. Here, $\vec{q}$ is one of the $N$ allowed phonon
wave vectors, while $j$ is the branch index, $j=1,2$, distinguishing
between the longitudinal and transversal branch. The normal
coordinates allow us to determine the normal mode bandstructure
$\lambda(\vec{q}j)$ through the relation
\begin{equation}\label{eq7}
 A(\vec{q}j) = \langle{}Q^*(\vec{q}j)Q(\vec{q}j)\rangle =
 \frac{k_{\mathrm{B}}T}{\lambda(\vec{q}j)}\:.
\end{equation}
Several experimental bandstructures of colloidal systems being based
on this relation can be found in literature
\cite{keim2004,danielprl,gomp2}. Starting from the overdamped Langevin
equation~(\ref{eq5}), one finds that the normal-mode auto-correlation
function decays exponentially \cite{hurd1982,baumgartl07},
\begin{equation}\label{eq6}
    \langle{}Q^*(\vec{q}j,t+\tau)Q(\vec{q}j,t)\rangle
=A(\vec{q}j)\exp\bigg(-\frac{\tau}
{T(\vec{q}j)}\bigg)
\end{equation}
with the overdamped phonon dispersion relation
\begin{equation}\label{eq8}
    T(\vec{q}j)=\frac{\gamma}{\lambda(\vec{q}j)}
\end{equation}
connecting the phonon decay times $T(\vec{q}j)$ with the elastic
bandstructure $\lambda(\vec{q}j)$. So far, the friction coefficient
$\gamma$ has been taken to be constant, relying on the assumption of
the Langevin model. Going beyond this model, this latter equation
takes a different form.  Hurd et al. \cite{hurd1982} showed that along
certain high symmetry directions (path sections 1 and 3, see central graph in top row in Fig.~\ref{fig2}),
hydrodynamic interactions lead to an overdamped phonon dispersion
relation of the more general form,
\begin{equation}\label{eq9}
    T(\vec{q}j)=\frac{\gamma(\vec{q}j)}{\lambda(\vec{q}j)}.
\end{equation}
where the friction coefficient $\gamma(\vec{q}j)$ is no longer
constant, but depends on the band index and the wave vector $\vec{q}$.

Having introduced these equations, we can formulate more precisely
the central idea of the present work. Equation~(\ref{eq7}) allows us
to determine the bandstructure by taking
$\lambda(\vec{q}j)=k_{\mathrm{B}}T/A(\vec{q}j)$. Equation~(\ref{eq6}), on the
other hand, permits an independent determination of the decay times
$T(\vec{q}j)$. Multiplication of both quantities according to
eq.~(\ref{eq9}) gives us access to the wave-vector dependent
friction coefficients,
\begin{equation}\label{eq10}
    \gamma(\vec{q}j)=\lambda(\vec{q}j)T(\vec{q}j)
\end{equation}
allowing a detailed experimental study of $\gamma(\vec{q}j)$ and thus
also a check of the Langevin model and its assumption of constant
friction.

The colloidal crystal dynamics can also be studied by examining the
behavior of the particle auto-correlation function
\begin{align}
\label{e49}
\nonumber
c(\tau)&= \frac{1}{N} \sum_{n,\alpha} \langle u_{n\alpha}(t+\tau)
u_{n\alpha}(t) \rangle \\
&= \frac{1}{N} \sum_{\vec{q},j} \langle
Q(\vec{q}j,t+\tau)Q^\ast(\vec{q}j,t) \rangle \:,
\end{align}
which is connected to the time-dependent mean-square displacement
$\delta_{\rm MSD}(\tau)$ through
\begin{equation}
\label{e51}
c(\tau)= \frac{1}{2}(\delta_{\rm MSD}(\infty)-\delta_{\rm
MSD}(\tau))\:.
\end{equation}
Within the Langevin description, this auto-correlation function is
related to the Laplace transform of the phonon spectrum
\cite{baumgartl07}
\begin{equation}\label{e52}
 \frac{k_{0}c(\tau')}{k_{\mathrm{B}}T}
= 2\int{}d\lambda' \frac{G(\lambda')}{\lambda'}\exp(-\lambda'\tau')
\end{equation}
where $\tau'= \tau k_0/\gamma$, $\lambda'=\lambda/k_0$ and where
\begin{equation}
\label{e39a}
G(\lambda) = \frac{1}{2N} \sum_{\vec{q}j} \delta(\lambda
- \lambda(\vec{q}j))\:,
\end{equation}
is the phonon spectrum, counting the number of $\lambda(\vec{q}j)$
within the interval $[\lambda,\lambda+d\lambda]$. This relation takes
a static quantity -- the system $\lambda(\vec{q}j)$ of
$\vec{q}$-dependent spring constants -- to predict a quantity
$c(\tau)$ that characterizes the dynamics of the system.  In classical
atomic crystals, a similar relation exists, connecting the Fourier
transform of the velocity auto-correlation function with the phonon
density of states \cite{dickey1969}. Equation~(\ref{e52}) is, of
course, valid only within the framework of the Langevin model.

\section{Results}
\label{sec:results}
\subsection{bandstructure, phonon decay and friction factors}
\label{sec:result1}

\begin{figure*}
\begin{center}
\includegraphics[width=0.7\textwidth]{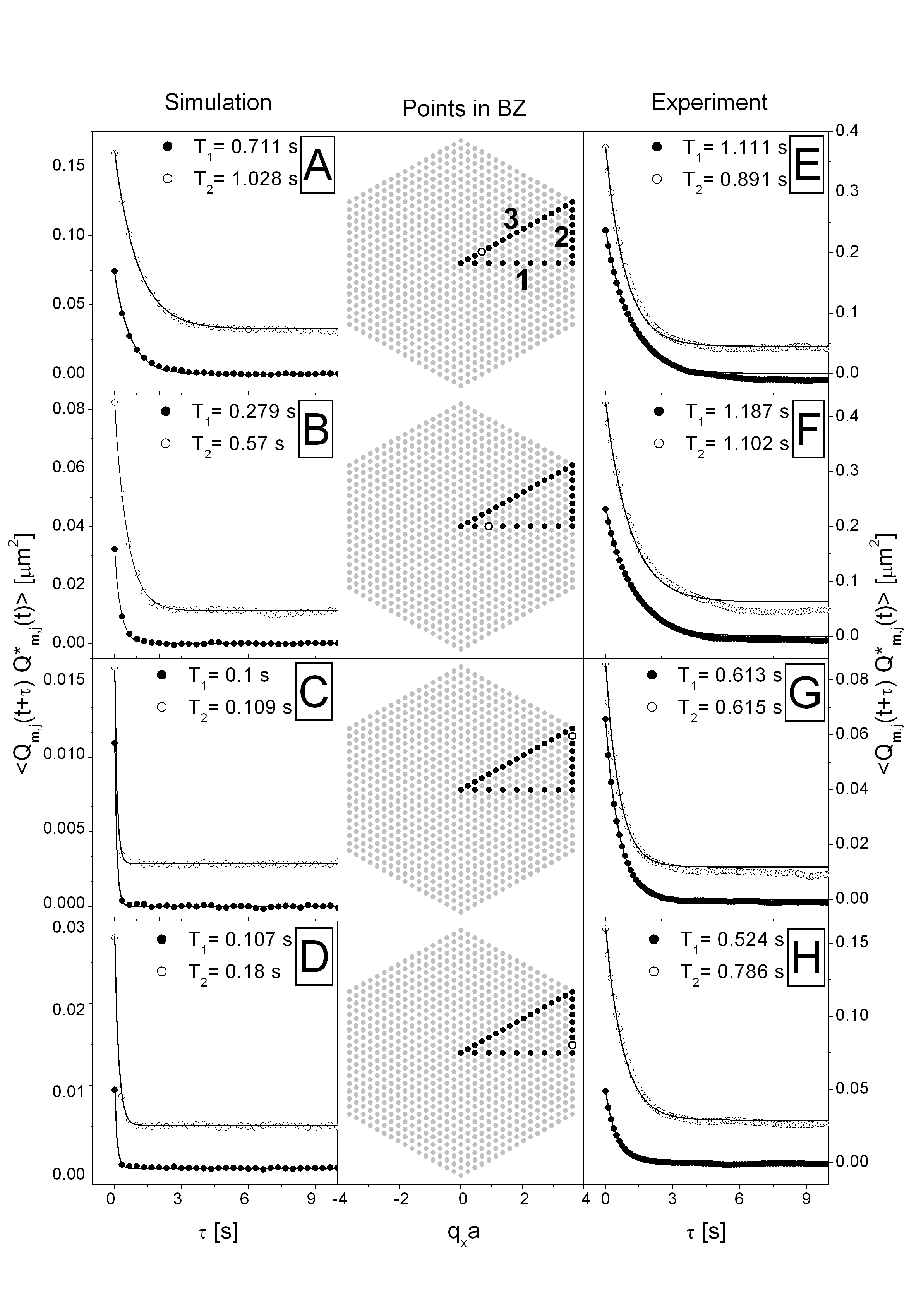}
\end{center}
\caption{Phonon auto-correlation function for the two band indices
$j=1$ (closed symbols) and $j=2$ (open symbols) at certain $\vec{q}$
values in the first Brilloun zone, as determined from Langevin
simulation data (A-D) and from video-microscopy data (E-H). The center column shows the 817 allowed phonon wave vectors
$\vec{q}$ in the first Brillouin zone; the $\vec{q}$ value actually
considered in the respective row is represented as a black open symbol, 
$\vec{q}$-vectors along the irreducible path are printed as solid black symbols, 
gray color is used otherwise. The numbers in the top graph label the sections of
the path around the irreducible part of the first Brillouin zone.
The solid lines are exponential fits to the data with the
resulting phonon decay times given in the legend of each figure. The
curves for $j=2$ (open symbols) are vertically shifted for clarity.}
\label{fig2}
\end{figure*}

\begin{figure}[t]
\begin{center}
\includegraphics[width=0.48\textwidth]{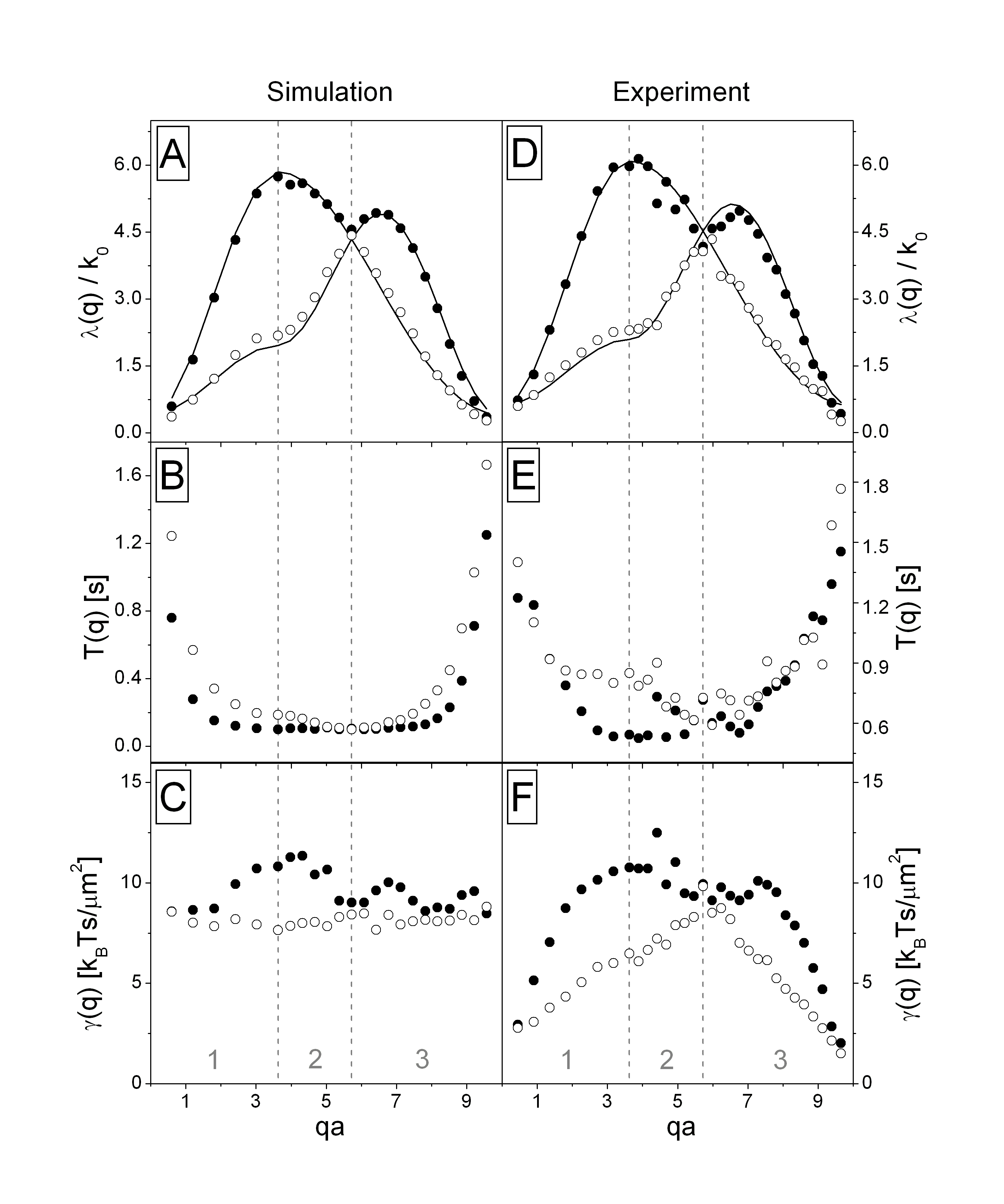}
\end{center}
\caption{Elastic bandstructure $\lambda(\vec{q}j)$ (top pair of
graphs), phonon decay times $T(\vec{q}j)$ (central pair of graphs),
and wavelength-dependent friction coefficient $\gamma(\vec{q}j)$ (bottom
row), plotted along the irreducible path 1 to 3 of the first Brillouin zone (see top row in Fig.~\ref{fig2}).  Solid lines in the
upper graph represent bandstructures calculated within the framework
of harmonic lattice theory.}
\label{fig3}
\end{figure}

\begin{figure}[t]
\begin{center}
\includegraphics[width=0.48\textwidth]{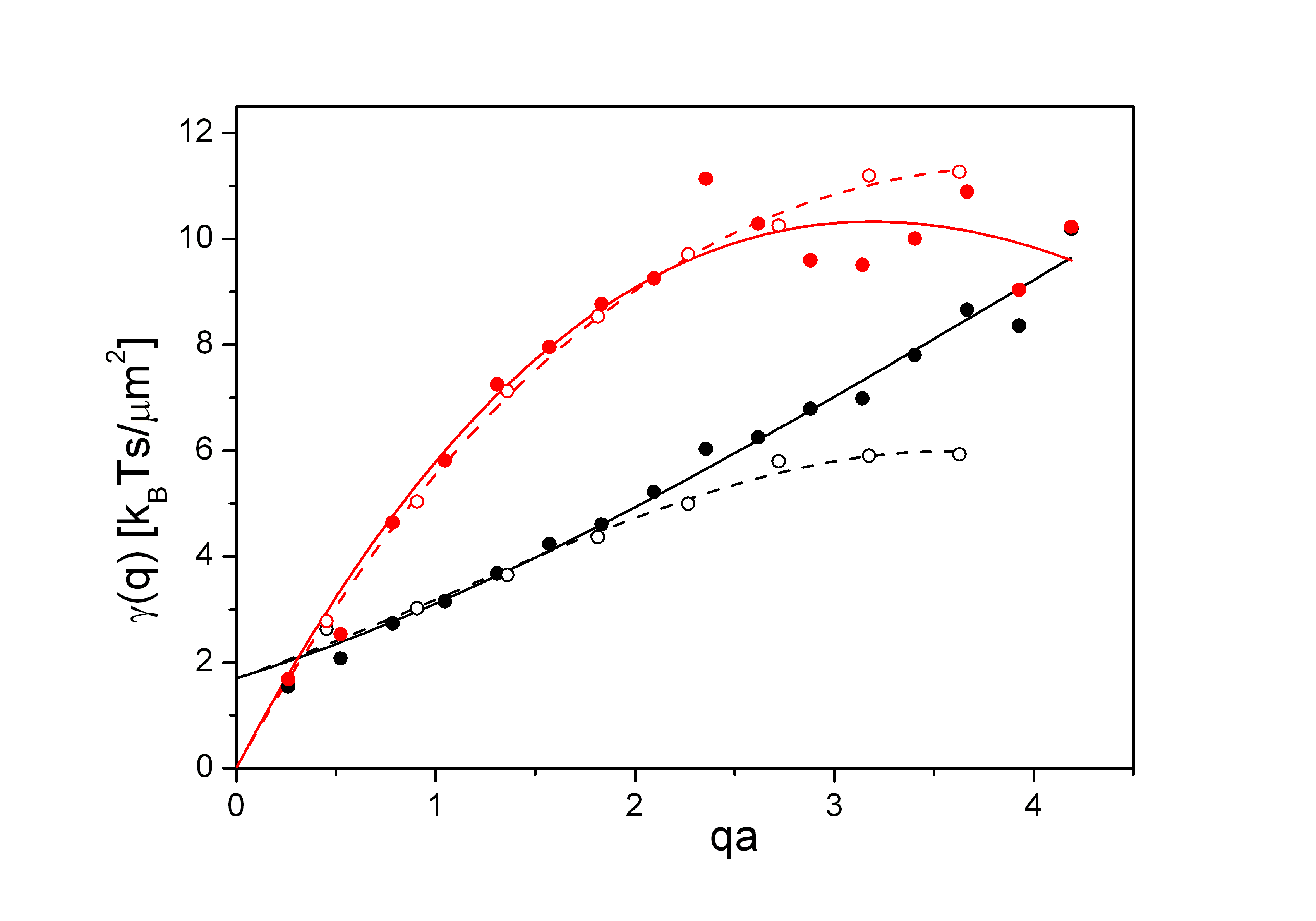}
\end{center}
\caption{Measured friction coefficients $\gamma(\vec{q}j)$ as shown in Fig.~\ref{fig3}F together with fitted polynomials 
(lines). Fit parameters are listed in table \ref{tab}. Upper two curves show the coefficients for the longitudinal
modes, while the lower two curves correspond to the transversal
modes. Open symbols for a direction in $\vec{q}$-space along path 1 (see top row
in Fig.~\ref{fig2}), closed symbols for direction along path 3.}
\label{fig4}
\end{figure}

\begin{table}[b]

\begin{center}

\begin{tabular}{l l r r r r}

\hline
&path&A&B&C&D \\ [0.5ex]
\hline
longitudinal &3&0&7.17&-1.45&0.07\\
longitudinal &1&0&6.72&-1.24&0.07\\
tranversal   &3&1.7&1.16&0.28&-0.02\\
tranversal   &1&1.7&1.28&0.29&-0.09
\end{tabular}
\caption{\label{tab}Fit-parameters of the function $f(x) = A + B x + C x^2 + D
  x^3$ used in Fig.~\ref{fig4}.}

\end{center}

\end{table}

\begin{figure}[t]
\begin{center}
\includegraphics[width=0.48\textwidth]{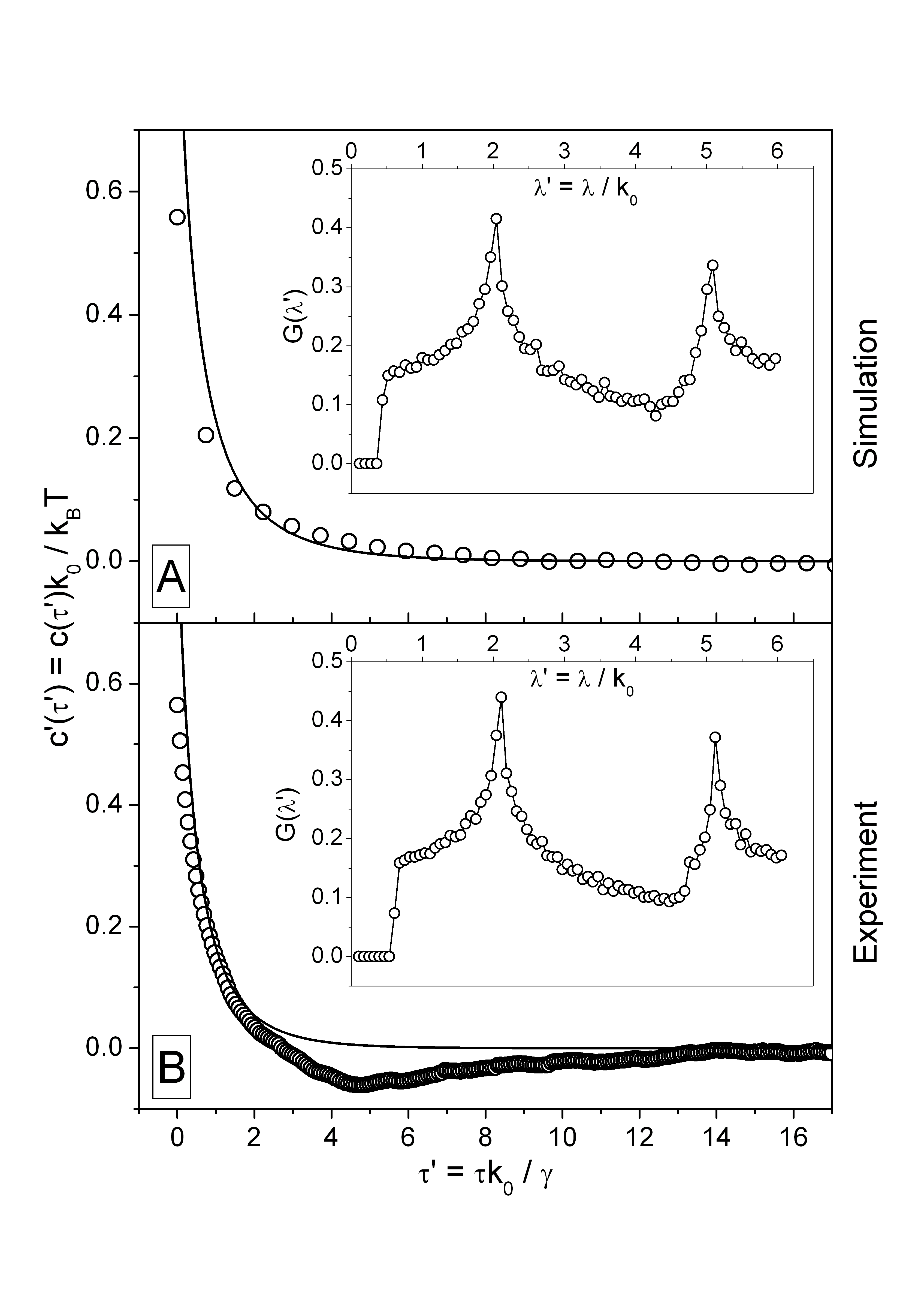}
\end{center}
\caption{Insets: phonon spectra, as obtained from
applying eq.~(\ref{e39a}) to the simulated and experimental
bandstructures $\lambda(\vec{q}j)$ presented in Fig.~\ref{fig3}. Main figures: particle
auto-correlation functions obtained from the
phonon spectra through a Laplace transformation (solid lines) and by
direct evaluation and averaging of $\langle u_{n\alpha}(t+\tau)
u_{n\alpha}(t) \rangle$ following eq.~(\ref{e49}) (open symbols).
A: simulation, B: experiment.}
\label{fig5}
\end{figure}

Figure~\ref{fig2} shows phonon auto-correlation functions computed
either from simulation data (left column of plots) or from
experimental configurations (right column of plots) for four different
allowed wave vectors $\vec{q}$ in the first Brillouin zone (center
column of plots). As outlined in eqs.~(\ref{eq6}) to (\ref{eq10}), we
expect an exponential decay and therefore fitted the data to the
function $A\exp(-\tau/T)$. While such an exponential decay is indeed
observed in the Langevin simulation data the experimental data show
some deviations at larger times. The experimental and simulation
curves are obtained from averages over 29500 and 10000 configurations,
respectively. Averaging over time windows of different lengths, one
can show that for correlation functions near to the edge of the first
Brillouin zone (BZ) these deviations of the experimental curves are due to insufficient
sampling. However, for correlation functions near the center of the
first BZ, one finds deviations which seem to be independent of the
average procedure. This might suggest that these (rather small)
deviations are fingerprints of propagating modes which are not
considered in our overdamped Langevin description and can thus not be
reproduced with exponential fit functions. Such propagating modes
are expected when the system is damped, but not overdamped, i.e., when
the damping is not strong enough to prevent an oscillatory
behavior. The alternative explanation -- that these deviations result
from the typical 2D instability against low $\vec{q}$ phonons -- can
be safely ruled out because the system is pinned to a hexagonal
lattice by the light-induced substrate.

Performing such fits for all allowed wave vectors along the irreducible path 1 to 3 (see top row in Fig.~\ref{fig2}), one obtains the function
$T(\vec{q}j)$, that is, the phonon decay times for both branches as a
function of the wave vector. These times are plotted in the central
row of Fig.~\ref{fig3}, for both the experimental as well as the
simulation data. On the other hand, one can use equation~(\ref{eq7})
to determine the elastic bandstructure, plotted for both sets of data
in the top row of Fig.~\ref{fig3}. Harmonic lattice theory of 2D
colloids in light-induced potentials \cite{baumgartl07} provides us
with explicit expressions for the bandstructure; they are fitted in
Fig.~\ref{fig3} to the measured structure using $k_0$ and $k_1$ as
fitting parameters. For the bandstructures produced from the
simulated data, we obtain $k_1/k_0 = 0.49$ and $k_0 = 18.9 \,k_{\mathrm{B}}T/
\upmu \mathrm{m}^2$
in good agreement with the values actually used in the simulation,
while from the experimental structures we derive $k_1/k_0 = 0.61$ and
$k_0 = 3.3\, k_{\mathrm{B}}T/\upmu \mathrm{m}^2$. With that we can estimate with eq.~(\ref{eq2})
(asumming $Z_{eff}=14000$) a value for $1/\kappa$ of $458\, \mathrm{nm}$ which is
actually quite a reasonable value.  As expected, due to the presence
of the stabilizing substrate the bandstructures are slightly shifted
upwards, with a shift that is directly proportional to the constant
$k_1$.

Following eq.~(\ref{eq10}) we multiply the two quantities
$\lambda(\vec{q}j)$ and $T(\vec{q}j)$ to obtain the friction
coefficients $\gamma(\vec{q}j)$, plotted in the bottom row of
Fig.~\ref{fig3}. In the simulations we used the value 
$\gamma=8.33 \,k_{\mathrm{B}}T \mathrm{s}/\upmu \mathrm{m}^2$ and a similar value 
$\gamma=8.85 \,k_{\mathrm{B}}T \mathrm{s}/\upmu \mathrm{m}^2$, independent of the wave-vector, comes out from the evaluation procedure.
This demonstrates the internal consistency of our scheme and
the correct implementation of our data evaluation procedures. More
importantly, it shows (i) that our statistics (which in the experiment
is even better) is sufficient, and (ii) that finite size effects are
negligible. While in the experiment we analysed a 16 shell hexagonal
crystallite being part of a much larger crystal, the simulation has
been carried out with 1000 particles and analysed within a 12 shell
hexagonal crystallite comprising 469 particles. Finite size effects
would show up in strong variations of $\gamma(\vec{q}j)$ near the
center of the BZ - something we do not observe in Fig.~\ref{fig3}.
Since the experimental system size is considerably larger than in the
simulation, we can be rather confident that also the experimental data
are not polluted by finite size effects.

With that we turn to our main result -- the function
$\gamma(\vec{q}j)$ derived from the experimental data (Fig.~\ref{fig3}F). We observe pronounced variations of the
friction coefficient, having roughly constant values only in one band
and only very close to the edge of the BZ. Most importantly, for both
bands $\gamma(\vec{q}j)$ tends to zero for $\vec{q} \to 0$. Almost the
same function $\gamma(\vec{q}j)$ is obtained when the renormalized simulated
bandstructure is multiplied with the experimental decay times. This
finding demonstrates that it is essentially the differences in
$T(\vec{q}j)$ which lead to the differences in $\gamma(\vec{q}j)$
between experiment and simulation. The different $\vec{q} \to 0$
behavior of the friction coefficients can then be traced back to the
decay times in this limit being considerably longer in the simulation
than in reality, implying that experimentally the long wavelength
phonons are decaying much faster than the Langevin simulation with its
constant friction assumption suggests. This is an effect due to
hydrodynamic interactions that shall be further discussed in the last
section.

Figure~\ref{fig4} shows again the measured friction factors: now the
coefficients for both $\vec{q}$ directions, path 1 and 3 (see top row in
Fig.~\ref{fig2}), appear over the same axis. Also, the coefficients
have been averaged over three equivalent directions in
$\vec{q}$-space. Note that consideration of all other directions is obsolete 
due to symmetry reasons.
To facilitate easy reproduction of our results we
have fitted the curves to a polynomial, with the fitting parameters
given in table~\ref{tab}. We observe for $q \to 0$ that while the
fitted functions of the longitudinal branches go to zero, those of the
two transversal modes extrapolate our results to a value slightly
larger than zero. We expect that this extrapolation is not very
meaningful as the real curve between $q=0$ and our first measured data
point might be completely different because of the emergence of
propagating phonons in this $\vec{q}$ regime which would destroy the
basis of our determination of $T(\vec{q}j)$. 

We finally observe that the theoretical Stokes friction coefficient $6 \uppi \eta (\sigma/2)$
for our colloidal spheres in water is 5 $k_{\mathrm{B}}T \mathrm{s}/\upmu \mathrm{m}^2$, showing that
our coefficients are of a reasonable magnitude (see Fig.~\ref{fig3} and Fig.~\ref{fig4}). This statement is supported by the observation presented in \cite{feitosa1991} where $\gamma$ increases to $8-10$ $k_{\mathrm{B}}T \mathrm{s}/\upmu \mathrm{m}^2$ if the colloidal sphere is located approximately 2 $\upmu \mathrm{m}$ above a wall. 

\subsection{Phonon spectrum and mean-square displacement}
\label{sec:result2}

Instead of considering the dynamics of all 2N phonons individually,
one may also study the colloidal dynamics more globally by examining
the average over all {\em phonon} auto-correlation functions $\langle
Q(\vec{q}j,t+\tau)Q^\ast(\vec{q}j,t) \rangle$, thus arriving at the
{\em particle} auto-correlation function $c(\tau)$ defined in
eq.~(\ref{e49}). As an average over all phonons, it is perhaps more
suited to serve as a general measure of the colloidal dynamics;
physically it is directly linked to the time dependent mean-square
displacement, eq.~(\ref{e51}).

To see how the observed differences between the actual colloid
dynamics and the Langevin dynamics manifest itself with respect to the
particle auto-correlation function $c(\tau)$, a Langevin prediction of
$c(\tau)$ for the experimental system is required.  Here, we can
exploit the relation (\ref{e52}) which assuming a Langevin system
derives a $c(\tau)$ directly from the phonon spectrum $G(\lambda)$ and
thus essentially from the bandstructure $\lambda(\vec{q}j)$. A
$c(\tau)$ thus predicted can then be confronted with the actual
particle auto-correlation function obtained from averaging $ \langle
u_{n\alpha}(t+\tau) u_{n\alpha}(t) \rangle$ as given in
eq.~(\ref{e49}). This is the idea of Fig.~\ref{fig5}.

The figure shows almost identical phonon spectra for both the
simulated and the experimental system. Several singularities can be
seen, the most important one is the jump singularity on the left edge
of the spectrum. If $k_1 \to 0$, i.e., if the substrate potential is
switched off, this singularity shifts to $\lambda=0$ and the phonon
spectrum $G(\lambda)$ takes a finite value at $\lambda=0$. This will
then lead to a divergent integral in eq.~(\ref{e52}) and, as a result,
$c(\tau)$ will logarithmically diverge \cite{baumgartl07}. This is the
well-known instability of 2D crystals which we here see to have been
cured thanks to the stabilizing effect of the substrate and the
light-induced shift of the jump singularity of $G(\lambda)$ away from
$\lambda=0$.

The Laplace transforms of the phonon spectra, eq.~(\ref{e52}), are
given as solid lines in the main figures of Fig.~\ref{fig5}, and are
compared to the $c(\tau)$ obtained from processing particle
displacements of 6000 (experiment) and 2000 (simulation) different
configurations containing 1017 (experiment) and 563 (simulation)
particles. For the simulation, being based exclusively on the Langevin
model, we {\em must} find perfect agreement between both curves,
provided our scheme is consistent and the data analysis tools are free
from flaws. And, indeed, we do find consistency of the data. As for
the experimental system agreement can only be expected as long as the
Langvin model applies. We observe a reasonable agreement for the first
five seconds (up to $\tau'=2$) and marked deviations towards negative
values between $\tau'=2$ and $\tau'=10$. These deviations are not
produced by poor statistics, or by the instability of 2D crystals
which we have supressed by the substrate, but are a real feature of the
system. Very likely, it is -- in a real-space presentation -- the
effect of the wave-vector dependence of the friction-coefficients.

\section{Discussion and concluding remarks}

The novel aspect of the present paper is the experimental
determination of the wavelength dependence of the friction coefficients in
2D colloidal crystals and, based on these data, a comparison of the
colloidal dynamics to the more simplified Langevin
dynamics. Determining the $\vec{q}$-dependence of friction coefficients
requires to measure two quantities independently, the elastic
bandstructure $\lambda(\vec{q}j)$ and the phonon decay times
$T(\vec{q}j)$. The friction coefficients then follow from multiplying
both quantities. Such a measurement is possible only with
video-microscopy data -- a technique whose most appealing property is
that it allows us to directly ``see'' the overdamped dynamics of the
colloids, in the form of the exponential decay of the
phonon auto-correlation function in Fig.~\ref{fig2}.

Our main finding is that in the long-wavelength limit, the friction
coefficient vanishes, i.e., $\gamma(\vec{q}j) \to 0$ for $q \to
0$. This is clearly the case for the longitudinal modes while the
extrapolated curves for the transversal modes show a small offset at
$q=0$ which however is probably not a meaningful extrapolation.  The
vanishing of friction of long wavelength phonons is a very reasonable
result. Hurd et al. pointed out that in 3D colloidal crystals back
flow will give an extra damping to longitudinal modes such that for $q
\to 0$ the damping is even larger than the Stokes friction
coefficient. Transverse modes, on the other hand, have been shown not
to be affected by a backflow damping; the flow induced by the
collective motions of the spheres now adds constructively, leading to
vanishing friction coefficients at long wavelengths. Quite the same
mechanism can be postulated to work for our 2D system: the transverse
modes in the limit $q \to 0$ will also induce constructive
interference of the flow, an overall flow field occurs going in the
same direction than the moving colloids and, as a result, the
colloidal particle show no longer motion relative to the surrounding
fluid, leading to a vanishing of the friction and thus a faster decay
of the overdamped phonons. For longitudinal modes, our 2D system
differs from the 3D system in that the 2D layer of particles is
coupled to a third spatial dimension into which the flow is free to
move. This third dimension prevents the back-flow damping, and a
favorable interference of the flow into the third dimension should
again be the reason for the vanishing of the friction coefficient for
$q \to 0$. However, a more detailed explanation -- as well as a
theoretical reproduction of $\gamma(\vec{q}j)$ in Fig.~\ref{fig3} --
has to wait for a proper hydrodynamic theory describing our system.
Such a theory must also account for hydrodynamic effects resulting
from the walls confining our systems \cite{bhat05}. Our results might
also be interesting to be reproduced applying more advanced simulation
techniques \cite{falck04} taking into account many-particle
hydrodynamics.

With these results in mind, it is clear that taking a constant
friction coefficient is a rather gross assumption of the overdamped
Langevin model. In addition, this model ignores that vanishing
friction coefficients in the long wavelength limit necessarily permits
phonons to start propagating (though being still damped, but not
overdamped). Systematic deviations from the exponentials in
Fig.~\ref{fig2} might be seen as a first indication for the
existence of such modes. However, it is also evident from
Fig.~\ref{fig2} that given these deviations indeed signalize the
onset of propagating phonons, they are still rather weak and far from
being able to dominate the overall dynamics of the system. The success
and failure of the overdamped Langevin model is best summarized by
Fig.~\ref{fig5} showing a particle auto-correlation function
$c(\tau)$ that agrees to the Langevin prediction only within the first
few seconds. Beyond that time, clear devations towards negative values
in $c(\tau)$ (anti-correlated behavior) can be observed, a feature
which according to our previous remarks must be due to the combined
effect of the wave-length dependence of the friction coefficient and,
possibly, propagating modes. Such a time-delayed anticorrelation has
also been observed in the two-particle experiments and has been
interpreted in terms of the standard Oseen tensor hydrodynamic
coupling \cite{meiners1999}. Zahn et al. also showed that the Oseen
term in 2D colloidal suspensions can lead to an increase of the
self-diffusion \cite{pepe}. Again: how these results connect to our
findings and how exactly hydrodynamics produces the anticorrelation,
can only be clarified with an elaborate hydrodynamical theory. It is
our hope that the present paper can stimulate the interest in such
theoretical work.

As well as motivating research on hydrodynamics our work might open novel perspectives in studying crystals which exhibit
non-overdamped particle dynamics. Such crystals can be already realized using a dusty plasma \cite{nunomura2000} where the micron-sized 
dust particles can be observed with videomicroscopy similar to colloidal crystals.

\section{Acknowledgements}
This project received financial support from the Austrian Science Foundation (FWF) under project title P18762. Jure Dobnikar wants 
to acknowledge the financial support of the Slovene Research Agency under the Grant P1-0055.

\end{document}